\documentclass[runningheads]{svmult}

\usepackage{makeidx}   
\usepackage{graphicx}  
\usepackage{subeqnar}  
\usepackage{multicol}  
\usepackage{physprbb}  
\makeindex             

\begin{document}
\title*{Towards the Event Horizon - High Resolution VLBI Imaging
of Nuclei of Active Galaxies}
%
%

\toctitle{Towards the Event Horizon - High Resolution VLBI Imaging
of Nuclei of Active Galaxies}

%
%
\titlerunning{Towards the Event Horizon}
%
\author{T.P. Krichbaum\inst{1}
\and D.A. Graham\inst{1}
\and A. Witzel\inst{1}
\and J.A. Zensus\inst{1}
\and A. Greve\inst{2}
\and M. Grewing\inst{2}
\and M. Bremer\inst{2}
\and S. Doeleman\inst{3}
\and R.B. Phillips\inst{3}
\and A.E.E. Rogers\inst{3}
\and H. Fagg\inst{4}
\and P. Strittmatter\inst{4}
\and L. Ziurys\inst{4}
}
\authorrunning{T.P. Krichbaum et al.}
%
%
\institute{Max-Planck-Institut f\"ur Radioastronomie, Bonn, Germany
\and Institut de Radioastronomie Millim\'etrique, Grenoble, France
\and MIT-Haystack Observatory, Westford, MA, USA
\and Steward Observatory, University of Arizona, Tucson, AZ, USA}

\maketitle              

~~\\
\noindent
{\large \bf 1. Introduction}\\
Very Long Baseline Interferometry at millimetre wavelengths (mm-VLBI)
allows to image compact galactic and extragalactic radio sources
with micro-arcsecond resolution, unreached by other astronomical observing techniques.
Future global VLBI at short millimetre wavelengths therefore should allow to map
the direct vicinity of the Super Massive Black Holes (SMBH) located at the centres
of nearby galaxies with a spatial resolution of only a few to a few ten gravitational radii.
With the reduced intrinsic self-absorption at these short wavelengths, mm-VLBI opens 
a direct view onto the often jet-producing "central engine".

Here we report on new developments in mm-VLBI, with emphasis on 
experiments performed at the highest frequencies possible to date.
We demonstrate that global VLBI at 150 and 230 GHz now is technically feasible
and yields source detections with an angular resolution as high as $25 - 30 \mu$as.
The combination of the existing with future telescopes (e.g. CARMA, ALMA, LMT, etc.) will
improve present day imaging capabilities by a large factor. Within the next
decade, one therefore could expect direct images of galactic and extragalactic 
(super massive) Black Holes and their emanating outflows.\\

\noindent
{\large \bf 2. Imaging the Jet Base of M87 with 20 ${\bf R_{\rm S}}$}\\
Since 2002, the Global mm-VLBI Array observes regularly at 86\,GHz (URL: www.mpifr-bonn.mpg.de/globalmm, cf.
Agudo et al., this conference). It combines the large European antennas (30\,m Pico Veleta,
6x15\,m Plateau de Bure, 100\,m Effelsberg, etc.) with the VLBA, and offers a factor
of $3-4$ higher sensitivity than the VLBA alone. As an example, we show in 
a new global-VLBI image of the inner jet of M87 at 86\,GHz (Figure 1). At a distance of 18.7 Mpc, 
the angular resolution of $300 \rm{x} 60$ $\mu$as corresponds to a spatial scale of 30 x 6 light days, or 
100 x 20 Schwarzschild-radii (assuming 3 x $10^6$ $\rm{M}_\odot$ for the SMBH). The 
existence of a fully developed jet on such small spatial scales gives important
new constraints for the theory of jet formation and may even indicate rotation of the 
central SMBH (via comparison with the width of the light cylinder).\\
\begin{figure}[t]
\includegraphics[bb=50 185 470 625,clip=,angle=-90,width=.4\textwidth]{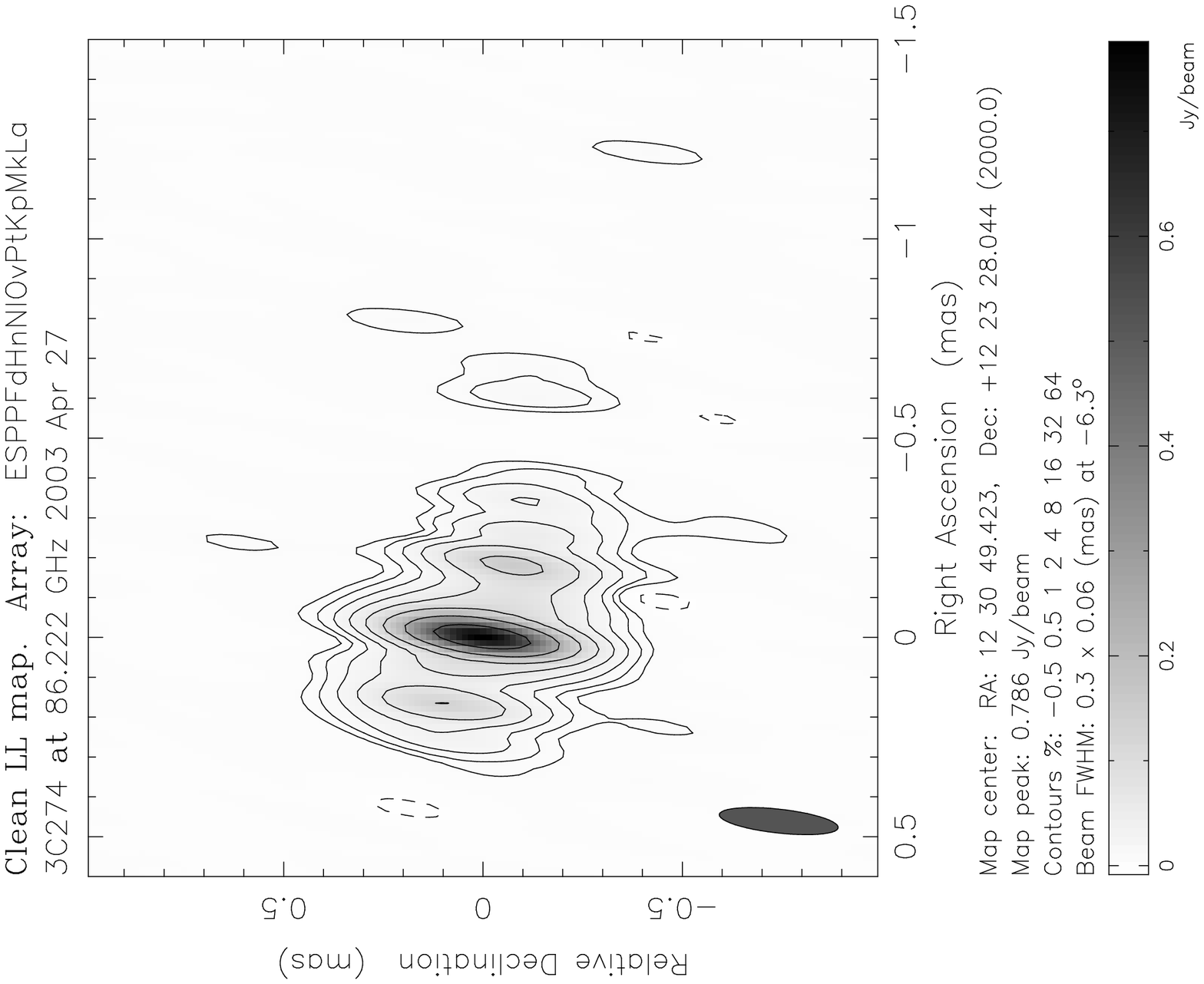}
\hspace{1cm}\begin{minipage}[t]{6cm}{
\vspace{0.3cm}
Figure 1: VLBI image of M87 (3C\,274) obtained in April 2003 at 86 GHz with the global mm-array.
Contour levels are -0.5, 0.5, 1, 2, 4, 8, 16, 32, and 64 \% of the peak flux of 0.79 Jy/beam. 
The beam size is 0.30 x 0.06 mas, pa=-6.3$^\circ$. The identification of the easternmost
jet component as VLBI core or as part of a counter-jet is still uncertain.}
\end{minipage}
\end{figure}

\noindent
{\large \bf 3. Towards Shorter Wavelengths - VLBI at 2 and 1\,mm}\\
A convincing demonstration of the feasibility of VLBI at wavelengths shorter than 3\,mm was made
at 2\,mm (147\,GHz) in 2001 and 2002. These first 2\,mm-VLBI experiments resulted in detections of 
about one dozen quasars on the short continental and long transatlantic baselines
(participating telescopes: Pico Veleta - Spain; Mets\"ahovi - Finland; Heinrich-Hertz and Kitt Peak
telescope - Arizona, USA). A big success was the
detection of 3 quasars on the 4.2\,G$\lambda$ long transatlantic baseline between Pico Veleta 
and the Heinrich-Hertz Telescope: NRAO150 (SNR=7), 1633+382 (SNR=23) and 3C279 (SNR=75). 
Motivated by this success, the observations were repeated in April 2003, this time at 1.3\,mm (230\,GHz).
Now also the phased IRAM interferometer on Plateau de Bure (France) participated. 
On the 1150\,km long baseline between Pico Veleta and Plateau de Bure the following sources were 
detected: NRAO\,150, 3C\,120, 0420-014, 0736+017, 0716+714, OJ287, 3C\,273, 3C\,279, and BL\,Lac.
On the 6.4\,G$\lambda$ long transatlantic baseline between Europe and Arizona fringes for
the quasar 3C454.3 (SNR=7.3) were clearly seen. For the BL\,Lac object 0716+714, however, only a
marginal detection (SNR=6.8) was obtained. These transatlantic detections
mark a new record in angular resolution in Astronomy (size $< 30 \mu$as). They indicate
the existence of ultra compact emission regions in AGN even at the highest frequencies 
(for 3C454.3 at z=0.859, the rest frame frequency is 428 GHz). So far, we find no evidence for
a reduced brightness temperature of the VLBI-cores at mm-wavelengths, however some variability is 
possible.\\

\noindent
{\large \bf 4. Future Outlook}\\
Good quality micro-arcsecond resolution VLBI images of the nuclei of galaxies will require an 
increased array sensitivity and better uv-coverage. The addition of large and sensitive mm-telescopes
like CARMA, ALMA, the LMT, etc. to the existing VLBI antennas will be crucial for the future success of
VLBI at and below 1\,mm.
The ongoing development towards observations with much larger larger bandwidths (several Gbits/s),
and for instantaneous atmospheric phase corrections and coherence prolongation (e.g. via water 
vapor radiometry), will further enhance the sensitivity. Thus one can hope that within less than a decade from now,
the detailed imaging of the `event horizon' of SMBHs and a better understanding of
the coupling between `central engine' and jet will become possible. 
\end{document}